\documentclass[useAMS,usenatbib]{mn2e}
\usepackage{graphicx}
\usepackage{txfonts}
\usepackage{natbib}
\usepackage{hyperref}
\usepackage{times,amssymb,amsfonts,amsbsy}

\begin{document}

\title[Shear peak counts]
  {Imprints of primordial non-Gaussianity on the number counts of cosmic shear peaks}

\author[M.\,Maturi et al.]{M.\, Maturi$^{1}$, C. Fedeli$^{2,3}$, and
  L. Moscardini$^{3,4,5}$\\
  $^1$ Zentrum f\"ur Astronomie, ITA, Universit\"at Heidelberg, Albert-\"Uberle-Str. 2, D-69120,
  Heidelberg, Germany\\
  $^2$ Department of Astronomy, University of Florida, 312 Bryant Space Science Center, Gainesville, FL 32611\\
  $^3$ Dipartimento di Astronomia, Universit\`a di Bologna, Via Ranzani 1, I-40127 Bologna, Italy \\
  $^4$ INFN, Sezione di Bologna, Viale Berti Pichat 6/2, I-40127 Bologna, Italy\\
  $^5$ INAF-Osservatorio Astronomico di Bologna, Via Ranzani 1, I-40127 Bologna, Italy}
      
\maketitle

\begin{abstract}
  We studied the effect of primordial non-Gaussianity with varied
  bispectrum shapes on the number counts of signal-to-noise peaks in
  wide field cosmic shear maps. The two cosmological contributions to
  this particular weak lensing statistic, namely the chance projection
  of Large Scale Structure and the occurrence of real, cluster-sized
  dark matter halos, have been modeled semi-analytically, thus
  allowing to easily introduce the effect of non-Gaussian initial
  conditions. We performed a Fisher matrix analysis by taking into
  account the full covariance of the peak counts in order to forecast
  the joint constraints on the level of primordial non-Gaussianity and
  the amplitude of the matter power spectrum that are expected by
  future wide field imaging surveys.  We find that positive-skewed
  non-Gaussianity increases the number counts of cosmic shear peaks,
  more so at high signal-to-noise values, where the signal is mostly
  dominated by massive clusters as expected. The increment is at the
  level of $\sim 1\%$ for $f_\mathrm{NL}=10$ and $\sim 10\%$ for
  $f_\mathrm{NL}=100$ for a local shape of the primordial bispectrum,
  while different bispectrum shapes give generically a smaller
  effect. For a future survey on the model of the proposed ESA space
  mission \emph{Euclid} and by avoiding the strong assumption of being
  capable to distinguish the weak lensing signal of galaxy clusters
  from chance projection of Large Scale Structures we forecasted a
  $1-\sigma$ error on the level of non-Gaussianity of $\sim 30-40$ for
  the local and equilateral models, and of $\sim 100-200$ for the less
  explored enfolded and orthogonal bispectrum shapes.
\end{abstract}

\begin{keywords} 
  cosmology: theory - gravitational lensing: weak - cosmological
  parameters - large-scale structure of the Universe
\end{keywords}

\section{Introduction}\label{sct:introduction}

Gravitational lensing is one of the most powerful means of
astrophysical investigation. While baryonic tracers of the dark matter
distribution commonly rely on strong simplifying assumptions, the
gravitational deflection of light \citep{BA01.1} is sensitive solely
to the overall incidence of any matter component along the line of
sight. In practical applications, the occurrence of strongly distorted
images in the core of massive galaxy clusters can return important
information about the inner structure of dark matter halos, while the
weak systematic image distortion of large numbers of background
galaxies allows to efficiently trace the outskirts of clusters and the
Large Scale Structure (LSS henceforth) in general.

Cosmic shear, that is weak gravitational lensing on cosmological
scales, measures the redshift evolution of the LSS in the Universe
weighted by a specific combination of angular diameter
distances. Hence it combines cosmological tests based on the geometry
of the Universe with tests based on the growth of
structures. Therefore, cosmic shear has long been recognized as an
important tool for cosmology. In this paper we considered one
particular cosmic shear statistic, namely the abundance of
signal-to-noise (S/N henceforth) ratio peaks in wide field weak
lensing maps. Besides the intrinsic ellipticity distribution of
background sources, peak counts are determined by the cosmological
information encapsulated by the cosmic matter density fluctuations
from linear to non-linear scales, namely LSS fluctuations and
cluster-sized dark matter halos. Counting cosmic shear peaks without
addressing their origin is not only easier to do than, for instance,
counting weak lensing selected galaxy clusters, but also contains more
cosmological leverage \citep{DI09.1}.

In the present work we applied the abundance of S/N peaks in cosmic
shear maps to the issue of cosmological initial conditions, and in
particular whether or not the primordial (dark) matter density
fluctuations are distributed according to a Gaussian distribution. It
is now well established that primordial non-Gaussianity has a
considerable impact on different aspects of structure
formation. Specifically, a positively (negatively) skewed distribution
of initial matter fluctuations would produce both a more (less)
abundant galaxy cluster population and more (less) biased structures
with respect to the underlying matter density field. Therefore, it is
expected that the peak statistics would return valuable constraints on
the level and shape of primordial non-Gaussianity.

Besides several pioneering works \citep{ME90.1,MO91.1,WE92.1}, the
problem of constraining deviations from primordial Gaussianity by
means different from the Cosmic Microwave Background (CMB) intrinsic
anisotropies has recently attracted renewed attention in the
literature, with efforts directed towards the abundance of non-linear
structures (\citealt*{MA00.2}; \citealt{VE00.1}; \citealt*{MA04.1};
\citealt{GR07.1,GR09.1,MA09.2}), halo biasing (\citealt{DA08.1,MC08.1};
\citealt*{FE09.1}; \citealt{FE10.2}), galaxy bispectrum
\citep{SE07.2,JE09.1}, mass density distribution \citep{GR08.2} and
topology \citep{MA03.2,HI08.2}, cosmic shear (\citealt*{FE10.1}, Pace
et al. 2010), integrated Sachs-Wolfe effect \citep*{AF08.1,CA08.1},
Ly$\alpha$ flux from low-density intergalactic medium \citep{VI09.1},
$21-$cm fluctuations \citep*{CO06.2,PI07.1} and reionization
\citep{CR09.1}.

As a specific application of our investigation on cosmic shear, we
refer to a future half-sky optical/near-infrared imaging survey on the
model of the ESA Cosmic Vision proposal \emph{Euclid} \citep{LA09.1}. Our
aim is to forecast the constraints on the level of primordial
non-Gaussianity that are expected by counting S/N peaks in cosmic
shear maps that should be produced by \emph{Euclid}. We shall also
investigate how these constraints change upon modification of several
survey parameters and detection criteria, such as the imaging depth,
the scale of optimal filtering, and the S/N threshold.

Throughout this work we adopted for the reference Gaussian model the
cosmological parameter set given by the latest analysis of the WMAP
data \citep{KO10.1}, namely the matter density parameter
$\Omega_{\mathrm{m},0}=0.272$, the cosmological constant density
parameter $\Omega_{\Lambda,0} = 0.728$, the baryon density parameter
$\Omega_{\mathrm{b},0}=0.046$, the Hubble constant $h\equiv
H_0/(100\,\mathrm{km\,s}^{-1}\,\mathrm{Mpc}^{-1})=0.704$, and the
matter power spectrum normalization set by $\sigma_8=0.809$. The rest
of our paper is organized as follows. In Section \ref{sct:ng} we
summarize the various non-Gaussian cosmologies adopted, while in
Section \ref{sct:observables} we recall the modifications that these
introduce to the cluster mass function, the large scale bias, and the
matter power spectrum. In Section \ref{sec:cosmolens} we describe the
various contributions to the weak lensing peak number counts and how
they have been modeled. In Section \ref{sct:effect} we show the effect
of primordial non-Gaussianity on the abundance of cosmic shear peaks,
while in Section \ref{sct:constraints} we report the Fisher matrix
analysis that we performed in order to forecast constraints on the
initial conditions. Finally in Section \ref{sct:conclusions} we
summarize our conclusions.

\section{Non-Gaussian cosmologies}\label{sct:ng}

Extensions of the most standard model of inflation
\citep{ST79.1,GU81.1,LI82.1} can produce substantial deviations from a
Gaussian distribution of primordial density and potential fluctuations
(see \citealt{BA04.1,CH10.1,DE10.1} for recent reviews). The amount
and shape of these deviations depend critically on the kind of
non-standard inflationary model that one has in mind, as will be
detailed later on.

A particularly convenient (although not unique) way to describe
generic deviations from a Gaussian distribution consists in writing
the gauge-invariant Bardeen's potential $\Phi$ as the sum of a
Gaussian random field and a quadratic correction
\citep{SA90.1,GA94.1,VE00.1,KO01.1}, according to
\begin{equation}\label{eqn:ng}
  \Phi = \Phi_\mathrm{G} + f_\mathrm{NL} * \left( \Phi_\mathrm{G}^2 -
  \langle \Phi_\mathrm{G}^2 \rangle \right) \;.
\end{equation}
On sub-horizon scales the Bardeen's potential equals minus the
Newtonian peculiar potential. The parameter $f_\mathrm{NL}$ in
Eq. (\ref{eqn:ng}) determines the amplitude of non-Gaussianity, and it
is in general dependent on the scale. The symbol $*$ denotes
convolution between functions, and reduces to standard multiplication
upon constancy of $f_\mathrm{NL}$. In the following we adopted the
large-scale structure convention (as opposed to the CMB convention,
see \citealt{AF08.1,CA08.1}; \citealt*{PI09.1} and \citealt{GR09.1})
for defining the fundamental parameter $f_\mathrm{NL}$. According to
this, the primordial value of $\Phi$ has to be linearly extrapolated
at $z = 0$, and as a consequence the constraints given on
$f_\mathrm{NL}$ by the CMB have to be raised by $\sim 30$ per cent to
comply with this paper's convention (see also \citealt*{FE09.1} for a
concise explanation).

In the case in which $f_\mathrm{NL} \ne 0$ the potential $\Phi$ is a
random field with a non-Gaussian probability distribution. Therefore,
the field itself cannot be described by the power spectrum
$P_\Phi({\boldsymbol k}) = Bk^{n-4}$ alone, rather higher-order
moments are needed, for instance the bispectrum $B_\Phi({\boldsymbol
  k}_1,{\boldsymbol k}_2,{\boldsymbol k}_3)$. The bispectrum is the
Fourier transform of the three-point correlation function $\langle
\Phi({\boldsymbol k}_1)\Phi({\boldsymbol k}_3)\Phi({\boldsymbol k}_3)
\rangle$ and it can hence be implicitly defined as
\begin{equation}
  \langle \Phi({\boldsymbol k}_1)\Phi({\boldsymbol
    k}_3)\Phi({\boldsymbol k}_3) \rangle \equiv
  (2\pi)^3\delta_\mathrm{D}\left( {\boldsymbol k}_1+{\boldsymbol
    k}_2+{\boldsymbol k}_3 \right) B_\Phi({\boldsymbol
    k}_1,{\boldsymbol k}_2,{\boldsymbol k}_3) \;.
\end{equation}

As mentioned above understanding the shape of non-Gaussianity is of
fundamental importance in order to pinpoint the physics of the early
Universe and the evolution of the inflaton field in particular. For
this reason, in this work we considered four different shapes of the
potential bispectrum, arising from different modifications of the
standard inflationary scenario. We summarize them in the following,
referring the reader to the quoted references and to \citet{FE10.2}
for further details.

\subsubsection*{Local shape}

The standard single-field inflationary scenario generates negligibly
small deviations from Gaussianity. These deviations are said to be of
the local shape, and the related bispectrum of the Bardeen's potential
is maximized for \emph{squeezed} configurations, where one of the
three wavevectors has much smaller magnitude than the other two. In
this case the parameter $f_\mathrm{NL}$ is usually assumed to be a
constant, and it is expected to be of the same order of the slow-roll
parameters \citep*{FA93.1}, that are very close to zero.

However non-Gaussianities of the local shape can also be generated in
the case in which an additional light scalar field, different from the
inflaton, contributes to the observed curvature perturbations
\citep*{BA04.2}. This happens, for instance, in curvaton models
\citep*{SA06.1,AS07.1} or in multi-fields models
\citep*{BA02.2,BE02.1}. In this case the parameter $f_\mathrm{NL}$ is
allowed to be substantially different from zero.

\subsubsection*{Equilateral shape}

In some inflationary models the kinetic term of the inflaton
Lagrangian is not standard, containing higher-order derivatives of the
field itself. One significant example of this is the DBI model
(\citealt*{AL04.1,SI04.1}, see also \citealt{AR04.1};
\citealt*{SE05.1,LI08.1}). In this case the primordial bispectrum is
maximized for configurations where the three wavevectors have
approximately the same amplitude, and it is well represented by the
template introduced by \citet{CR07.1}.

Given this, we have the freedom to insert a \emph{running}
$\gamma({\boldsymbol k}_1,{\boldsymbol k}_2,{\boldsymbol k}_3)$ for
$f_\mathrm{NL}$, since this parameter is not forced to be constant in
this case. The form we chose reads \citep{CH05.2,LO08.1,CR09.1}
\begin{equation}
  \gamma({\boldsymbol k}_1,{\boldsymbol k}_2,{\boldsymbol k}_3) = \left(
  \frac{k_1+k_2+k_3}{k_\mathrm{CMB}} \right)^{-2\kappa} \;.
\end{equation}
In all calculations that follow we considered this running as part of
the equilateral bispectrum. This is important since different authors
choose different runnings, or no running at all, for the equilateral
shape. We adopted the exponent $\kappa=-0.2$ in the remainder of this
work, that increase the level of non-Gaussianity at scales smaller
than that corresponding to $k_\mathrm{CMB} = 0.086 h$ Mpc$^{-1}$. This
coincides with the larger multipole used in the CMB analysis by the
WMAP team \citep{KO09.1,KO10.1}, $\ell\sim 700$.

\subsubsection*{Enfolded shape}

For deviations from Gaussianity evaluated in the regular Bunch-Davies
vacuum state, the primordial potential bispectrum is of local or
equilateral shape, depending on whether or not higher-order
derivatives play a significant role in the evolution of the inflaton
field. If the Bunch-Davies vacuum hypothesis is dropped, the resulting
bispectrum is maximal for \emph{squashed} configurations
\citep{CH07.1,HO08.1}. \cite*{ME09.1} found a template that describes
very well the properties of this enfolded-shape bispectrum (see
however \citealt{CR10.1} for a slightly different template of the
physical model).

\subsubsection*{Orthogonal shape}

A shape of the bispectrum can be constructed that is nearly
\emph{orthogonal} to both the local and equilateral forms
\citep*{SE10.1}. Constraints on the level of non-Gaussianity
compatible with the CMB in the local, equilateral and orthogonal
scenarios were recently given by the WMAP team \citep{KO10.1}, while
constraints on enfolded non-Gaussianity from galaxy bias
were given by \citet{VE09.1}\\
\\\indent Although there is no theoretical prescription against a
running of the $f_\mathrm{NL}$ parameter with the scale in the
enfolded and orthogonal shapes, we decided not to include one. The
reason for this is that there is no first principle that can guide one
in the choice of a particular kind of running, and until now no work
has addressed the problem of a running for these shapes
\citep*{FE09.2,FE09.3}. Moreover, we recall that the four non-Gaussian
shapes described above are not independent, rather the orthogonal
shape can be obtained as a suitable linear combination of the
equilateral and enfolded shapes \citep*{SE10.1,WA10.1}. Nevertheless,
we performed computations for the orthogonal model as well, since it
gives a different signature on the evolution of the LSS.

\section{Cosmological observables}\label{sct:observables}

Primordial non-Gaussianity produces modifications in the statistics of
density peaks, resulting in differences in the mass function of cosmic
objects  and  the  bias of  dark  matter  halos  with respect  to  the
underlying  smooth density field.  In the  following we  summarize how
these modifications have been taken  into account in the present work,
and how they reflect into modifications to the matter power spectrum.

\subsection{Mass function}\label{sct:mf}

For the non-Gaussian modification to the mass function of cosmic
objects we adopted the prescription of \cite{LO08.1}. The main
assumption behind it is that the effect of primordial non-Gaussianity
on the mass function is independent of the prescription adopted to
describe the mass function itself. This means that, if
$n^\mathrm{(G)}_\mathrm{PS}(M,z)$ and $n_\mathrm{PS}(M,z)$ are the
non-Gaussian and Gaussian mass functions, respectively, computed
according to the \cite{PR74.1} formula, we can define a correction
factor $\mathcal{R}(M,z)\equiv
n^\mathrm{(G)}_\mathrm{PS}(M,z)/n_\mathrm{PS}(M,z)$. Then, the
non-Gaussian mass function computed according to an arbitrary
prescription, $n(M,z)$ can be related to its Gaussian counterpart
through
\begin{equation}
  n(M,z) = \mathcal{R}(M,z) n^\mathrm{(G)}(M,z) \;.
\end{equation}

In order to compute $n_\mathrm{PS}(M,z)$, and hence
$\mathcal{R}(M,z)$, \citet{LO08.1} performed an Edgeworth expansion
\citep{BL98.1} of the probability distribution for the smoothed
density fluctuations field, truncating it at the linear term in
$\sigma_M$. The resulting \cite{PR74.1}-like mass function reads
\begin{eqnarray}\label{eqn:mfps}
n_\mathrm{PS}(M,z) &=& - \sqrt{\frac{2}{\pi}}
\frac{\rho_\mathrm{m}}{M} \exp\left[
  -\frac{\delta_\mathrm{c}^2(z)}{2\sigma_M^2} \right] \left[
  \frac{d\ln \sigma_M}{dM} \left(
  \frac{\delta_\mathrm{c}(z)}{\sigma_M} + \right.\right.
  \nonumber\\ &+& \left. \left. \frac{S_3\sigma_M}{6} \left(
  \frac{\delta_\mathrm{c}^4(z)}{\sigma^4_M}
  -2\frac{\delta^2_\mathrm{c}(z)}{\sigma^2_M} -1\right) \right) +
  \right.  \nonumber\\ &+& \left. \frac{1}{6} \frac{dS_3}{dM}\sigma_M
  \left( \frac{\delta^2_\mathrm{c}(z)}{\sigma^2_M} -1\right) \right] \;.
\end{eqnarray}
In the previous equation
$\rho_\mathrm{m}=3H_0^2\Omega_{\mathrm{m},0}/8\pi G$ is the comoving
matter density in the Universe, $\sigma_M$ is the \emph{rms} of
density fluctuations smoothed on a scale corresponding to the mass
$M$, and $\delta_\mathrm{c}(z) = \Delta_\mathrm{c}/D_+(z)$. The
function $S_3(M)\equiv \mu_3(M)/\sigma_M^4$ is the reduced skewness of
the non-Gaussian distribution, and the skewness $\mu_3(M)$ can be
computed as
\begin{eqnarray}
  \mu_3(M) &=& \int_{\mathbb{R}^9} \mathcal{M}_R(k_1)
  \mathcal{M}_R(k_2) \mathcal{M}_R(k_3) \times
  \nonumber\\ &\times&\langle\Phi({\bf k}_1)\Phi({\bf k}_2)\Phi({\bf
    k}_3)\rangle \frac{d{\bf k}_1d{\bf k}_2d{\bf k}_3}{(2\pi)^9} \;.
\end{eqnarray}
The last thing that remains to be defined is the function
$\mathcal{M}_R(k)$, that relates the density fluctuations smoothed on
some scale $R$ to the respective peculiar potential,
\begin{equation}
  \mathcal{M}_R(k) \equiv
  \frac{2}{3}\frac{T(k)k^2}{H_0^2\Omega_{\mathrm{m},0}}W_R(k) \;,
\end{equation}
where $T(k)$ is the matter transfer function and $W_R(k)$ is the
Fourier transform of the top-hat window function.

In this work we adopted the \cite{BA86.1} matter transfer function,
with the shape factor correction of \cite{SU95.1}. This reproduces
fairly well the more sophisticated recipe of \citet{EI98.1} except for
the presence of the baryon acoustic oscillation, that anyway is not of
interest here. We additionally adopted as reference mass function the
prescription of \citet{SH02.1} (see \citealt{JE01.1,WA06.1,TI08.1} for
alternative prescriptions). Other approaches also exist for computing
the non-Gaussian correction to the mass function, that give results in
broad agreement with those obtained here \citep*{MA00.2}. In computing
the non-Gaussian corrections to the mass function we have taken into
account the correction to the critical overdensity for collapse
suggested by \citet{GR09.1} (see also
\citealt*{MA09.3,MA09.2,MA10.1}), according to which
$\Delta_\mathrm{c} \rightarrow \Delta_\mathrm{c}\sqrt{q}$, with $q\sim
0.8$.

\subsection{Halo bias}\label{sct:bs}

Recently much attention has been devoted to the effect of primordial
non-Gaussianity on halo bias, and the use thereof for constraining
$f_\mathrm{NL}$ (\citealt{MA08.1,DA08.1,VE09.1}; \citealt*{CA08.1}). In
particular, these works have shown that primordial non-Gaussianity
introduces a scale dependence on the large scale halo bias. This
peculiarity allows to place already stringent constraints from
existing data \citep{SL08.1,AF08.1}.

The non-Gaussian halo bias can be written in a relatively
straightforward way in terms of its Gaussian counterpart as
\citep*{CA10.1}
\begin{equation}
  b(M,z,k) = b^\mathrm{(G)}(M,z) + \beta_R(k)\sigma_M^2\left[
    b^\mathrm{(G)}(M,z)-1 \right]^2 \;,
\end{equation}
where the function $\beta_R(k)$ encapsulates all the scale dependence
of the non-Gaussian correction to the bias, and reads
\begin{eqnarray}
  \beta_R(k) &=& \frac{1}{8\pi^2\sigma_M^2\mathcal{M}_R(k)}
  \int_0^{+\infty} \zeta^2\mathcal{M}_R(\zeta) \times
  \nonumber\\ &\times& \left[ \int_{-1}^1
    \mathcal{M}_R\left(\sqrt{\alpha}\right) \frac{B_\Phi\left(
      \zeta,\sqrt{\alpha},k \right)}{P_\Phi(k)} d\mu \right] d\zeta \;,
\end{eqnarray}
where $\alpha \equiv k^2 + \zeta^2 + 2 k\zeta\mu$. In the simple case
of local bispectrum shape it can be shown that the function
$\beta_R(k)$ should scale as $\propto k^{-2}$ at large scales, so that
a substantial boost (if $f_\mathrm{NL}>0$) in the halo bias is
expected at those scales. For the Gaussian bias $b^\mathrm{(G)}(M,z)$
we adopted the prescription of \citet*{SH01.1}. In this case, since
the correction to the Gaussian bias is written in term of the Gaussian
bias itself, the ellipsoidal collapse correction suggested by
\citet{GR09.1} is not necessary.

It is interesting to note that, while for the first three non-Gaussian
shapes introduced in Section \ref{sct:ng}, a positive $f_\mathrm{NL}$
implies both a positive skewness of the matter density field and a
positive correction to the large scale halo bias, the opposite is true
for the fourth shape, the orthogonal one. This is a fact to be kept in
mind when interpreting our results in the subsequent Sections.

\subsection {Matter power spectrum}\label{sec:matterPWS}

Another cosmological observable that is relevant for our purposes and
gets modified in case of primordial non-Gaussianity is the matter
power spectrum. The latter requires a little bit more of care with
respect to the mass function and linear bias. Particularly, we should
have a reliable way to parametrize the fully non-linear
three-dimensional power spectrum, since part of the lensing signal is
given by the integral thereof along the line of sight. In order to do
that we followed the approach of \citet*{FE10.1} (and references
therein) and made use of the halo model in order to represent the
matter power spectrum.

The halo model \citep*{SE00.1,MA00.3,CO02.2} is a physically motivated
framework that allows to compute the correlation function of various
LSS tracers, including the dark matter particles themselves. It is
based on the simple consideration that if all the matter in the
Universe is locked into halos, then the contribution to the
correlation function comes from particle pairs sitting in separated
halos at large scales and from particle pairs belonging to the same
halo at small scales. Accordingly, the power spectrum can be written
as the sum of two terms describing the two contributions, $P(k,z) =
P_1(k,z) + P_2(k,z)$, with
\begin{equation}
  P_1(k,z) = \int_0^{+\infty} n(M,z) \left[
    \frac{\hat{\rho}(k,M,z)}{\rho_\mathrm{m}} \right]^2 dM \;,
\end{equation}
and
\begin{equation}
  P_2(k,z) = \left[\int_0^{+\infty} n(M,z)
    b(M,z,k)\frac{\hat{\rho}(k,M,z)}{\rho_\mathrm{m}} dM\right]^2
  P_\mathrm{L}(k,z) \;.
\end{equation}
In the previous set of equations $P_\mathrm{L}(k,z)$ is the linear
matter power spectrum and $\hat{\rho}(k,M,z)$ is the Fourier transform
of the mean dark matter halo density profile which is also in
principle modified by primordial non-Gaussianity, as suggested in
\citet{AV03.1} and \citet*{SM10.1}. Nevertheless, we adopted a
standard \citet*{NA96.1,NA97.1} shape (NFW henceforth) for
$\hat{\rho}(k,M,z)$ because the functional forms for the modified
density profiles are proven to hold only for the local shape, and it
is not clear whether this is the case for other shapes as
well. Therefore we ignored these changes, making our analysis more
conservative, since a positive (negative) $f_\mathrm{NL}$ implies a
more (less) centrally concentrated dark matter profile, and
consequently an increment (decrement) in the number of peaks.

For practical details on the implementation of the halo model we refer
the interested reader to \citet{AM04.1,FE10.1}. Here we just note that
the mass function and the halo bias, both entering in the halo model,
are modified by primordial non-Gaussianity according to Sections
\ref{sct:mf} and \ref{sct:bs}.

\section{Matter fluctuations with cosmic shear}\label{sec:cosmolens}

In this paper we focused on the number counts of weak lensing peaks,
that directly reflect the distribution of dark matter
fluctuations. Counting directly the peaks allows to minimize the
physical assumptions necessary in going from the adopted cosmological
model to the data outcome. As a matter of fact we based the final
results on dark matter physics only, by observing quantities
insensitive to baryon physics and by relating the model prediction
directly to the measured S/N ratios. In particular, the latter point
avoids the difficult task of disentangling in a clear way the
intrinsic nature of each single peak: noise fluctuation, LSS
line-of-sight superimposition or actual galaxy cluster.

\subsection {Lensing formalism}

The deflection properties of isolated lenses are fully described by
their two-dimensional lensing potential,
\begin{equation}\label{eq:l_potential}
  \psi( \boldsymbol{\theta}) \equiv \frac{2}{c^2}\frac{D_{\rm
      ds}}{D_{\rm d}D_{\rm s}}\int_0^s\Phi(D_{\rm d}\boldsymbol{\theta},
  z) dz \;,
\end{equation}
where $\Phi$ is the Newtonian gravitational potential and
$D_\mathrm{s}$, $D_\mathrm{d}$, and $D_\mathrm{ds}$ are the
angular-diameter distances between the observer and the source, the
observer and the lens, and the lens and the source,
respectively. Finally, $s$ represents the physical distance out to the
source sphere.

The potential $\psi$ relates the angular positions $\boldsymbol\beta$
of a source and $\boldsymbol\theta$ of its images on the observer's
sky through the lens equation,
$\boldsymbol{\beta}=\boldsymbol{\theta}-\nabla\psi$. For sources such
as distant background galaxies it is possible to linearize the lens
equation such that the induced image distortion is expressed by the
Jacobian matrix
\begin{equation}
  A = (1-\kappa)
  \left(
  \begin{array}{cc}
    1-g_1 & -g_2 \\
    -g_2 & 1+g_1 \\
  \end{array}
  \right) \;.
\end{equation} 
In the previous equation, $\kappa \equiv \nabla^2\psi/2$ is the
convergence, responsible for the isotropic magnification of an image
relative to its source, and $g=\gamma/(1-\kappa)$ is the reduced
shear, quantifying the observed distortion. Here,
$\gamma_1\equiv\left(\psi_{,11}-\psi_{,22}\right)/2$ and
$\gamma_2\equiv\psi_{,12}$ are the two components of the complex
shear, where a comma denotes standard differentiation with respect to
coordinates on the lens plane. It is important to recall that since
the angular size of the sources is unknown, only the reduced shear can
be estimated starting from the observed ellipticity of the images.

\subsection {Contributions to the lensing signal}\label{sec:cosmoConstraints}

The abundance of S/N peaks in cosmic shear maps is determined by the
sum of three separate contributions, that we describe in the
following.

\begin{enumerate}
\item The signal due to the occurrence of real non linear structures
  such as galaxy clusters. Their abundance is evaluated in
  Section~\ref{sct:mf} and their shear profile can be evaluated
  analytically (see \citealt{BA96.1,ME03.1}). Since the typical
  separation between the observed background galaxies is larger than
  the typical radius of the galaxy clusters' critical curves we can
  assume $\gamma\simeq g$ throughout.

\item The lensing signal due to chance projections of the LSS, given
  by the effective convergence power spectrum
  \begin{equation} 
    P_\kappa(\ell)=\frac{9H_0^4\Omega_{\mathrm{m},0}^2}{4
      c^4}\int_0^{w_{\rm
        H}}dw\frac{\bar{W}^2(w)}{a^2(w)}P\left(\frac{\ell}{f_K(w)},w\right) \;,
    \label{eq:lss} 
  \end{equation}
  where $P(k,z)$ is the fully non-linear matter power spectrum
  discussed in Section \ref{sec:matterPWS}, $\Omega_{\mathrm{m},0}$ is
  the present-day matter-density parameter, $H_0$ is the Hubble
  constant, $c$ is the speed of light, $a(w)$ is the scale factor at
  the comoving distance $w$, $w_{\rm H}$ is the comoving distance to
  the horizon, $f_K$ is the comoving angular-diameter distance, and
  $\bar W(w)$ is the weight function encorporating the line-of-sight
  integral over the distribution of background sources $G(w)$,
  \begin{equation} 
    \bar W(w)=\int_w^{w_{\rm H}}dw' G(w')\frac{f_K(w'-w)}{f_K(w')} \;.
  \end{equation}
  In our application we used a single (the tangential) component of
  the shear, for which the power spectrum reads
  $P_{\gamma_\mathrm{t}}(\ell)=P_\kappa(\ell)$/2.

\item The observational noise contribution from the intrinsic
  ellipticity and finite number of background galaxies used to measure
  the shear signal, which has the white noise power spectrum
  \begin{equation}
    P_\epsilon(\ell)=\frac{1}{2} \frac{\sigma_{\epsilon_\mathrm{s}}^2}{n_\mathrm{g}} \;,
    \label{eq:noisePower} 
  \end{equation}
  determined by the number density $n_\mathrm{g}$ of galaxies suitable
  for weak lensing measurements and the variance
  $\sigma_{\epsilon_\mathrm{s}}^2$ of their intrinsic ellipticity
  distribution.
\end{enumerate}

Since contributions (ii) and (iii) are well represented by two
independent Gaussian random fields, their total power spectrum, is the
sum of the two contributions,
$P(\ell)=P_{\gamma_\mathrm{t}}(\ell)+P_\epsilon(\ell)$.

\subsection {Optimal lensing filtering}\label{sec:filter}

\begin{figure*}
  \centering
  \includegraphics[width=0.5\hsize]{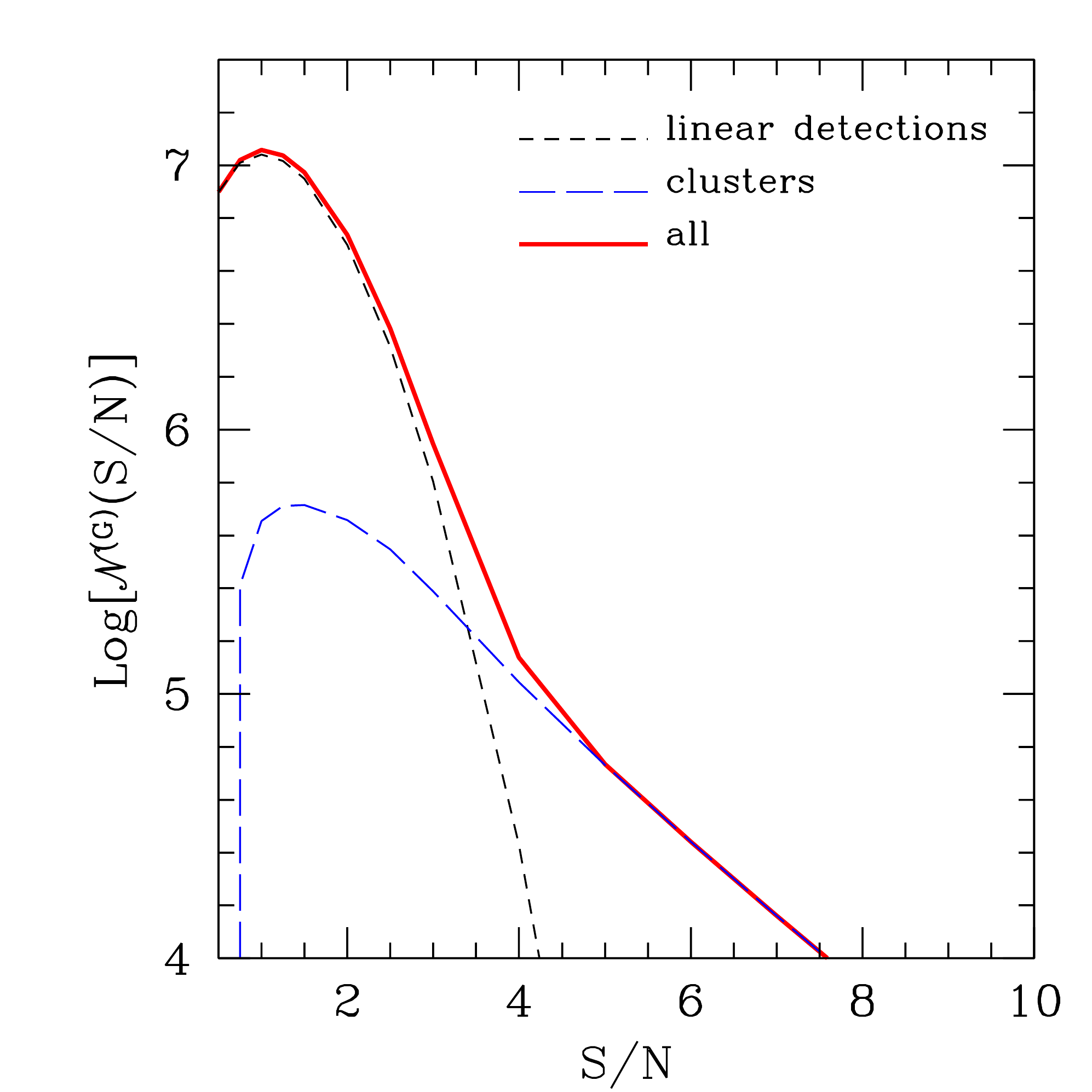}\hfill
  \includegraphics[width=0.5\hsize]{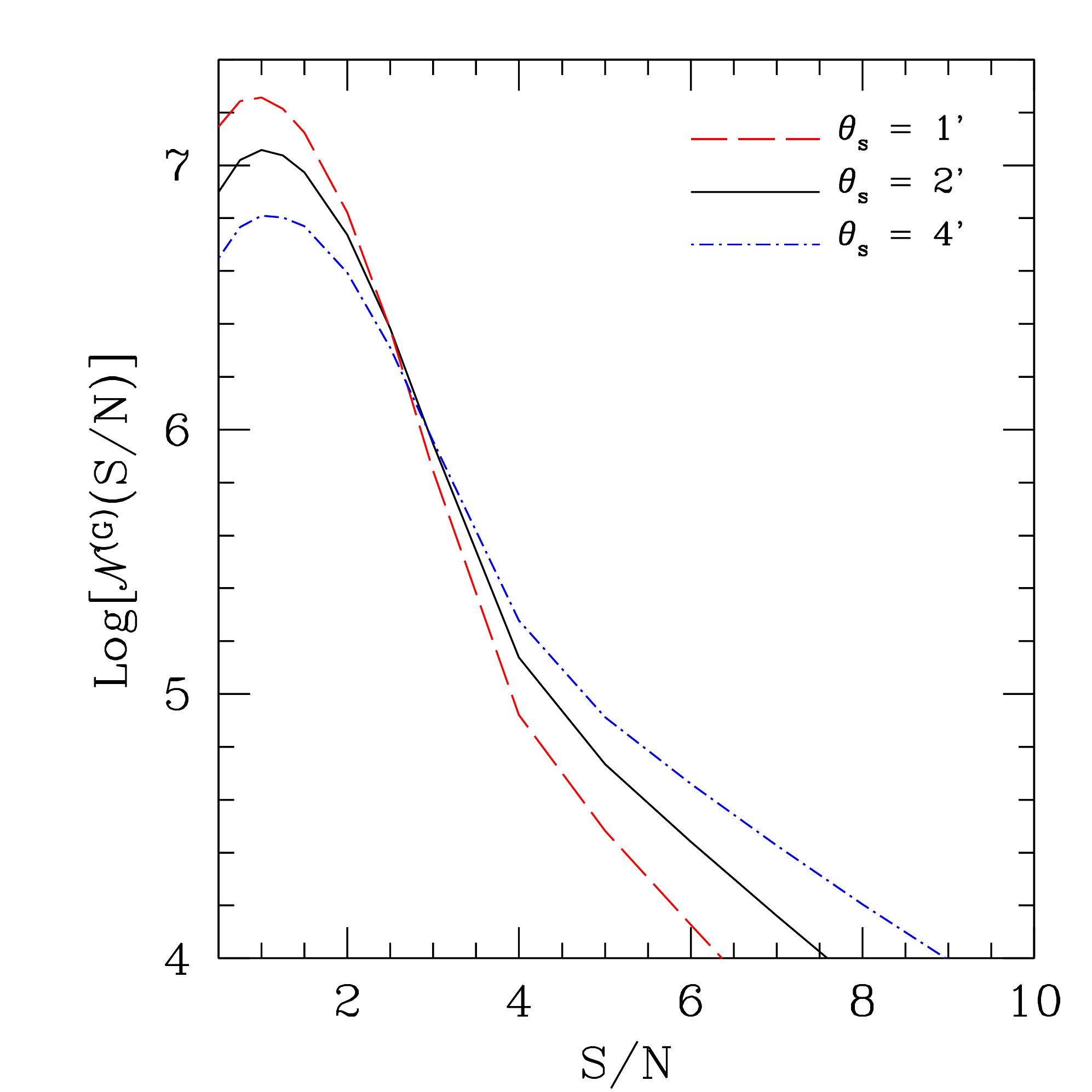}
  \caption{\emph{Left panel}. The different contributions to the
    abundance of shear S/N peaks in the reference Gaussian model. The
    black short dashed line shows the contribution given by LSS
    projection and data noise, while the blue long dashed line
    refers to the contribution of the real dark matter clumps. The
    heavy solid red line is the total count distribution. We adopted
    optimal filtering with a scale radius of $\theta_\mathrm{s} =
    2'$. \emph{Right panel}. The total number of shear S/N peaks for
    three different filter scale radii, as labeled.}
  \label{fig:shearPeaks}
\end{figure*}

A method to measure gravitational lensing signatures in
optical/near-infrared catalogues of galaxies is the aperture mass
\citep{SC98.1}. It is a weighted average of the tangential component
of the shear $\gamma_{\rm t}$ over the position $\boldsymbol{\theta}$
on the sky,
\begin{equation}
  \label{eq:A}
  A(\boldsymbol\theta)=\int_{\mathbb{R}^2}
  d^2\boldsymbol\theta'\gamma_{\rm
    t}(\boldsymbol{\theta}',\boldsymbol{\theta})
  Q(\|\boldsymbol{\theta}'-\boldsymbol{\theta}\|) \;,
\end{equation}

where the radial filter function $Q$ determines the statistical
properties of the estimate $A$ we are interested in. In this paper we
aim at constraining the deviations from primordial Gaussianity, which
mostly affect the non-linear part of structure formation and thus the
high mass end of the mass function. Therefore, we adopted the linear
matched filter defined by \citet{MA05.1} which is designed to maximize
the S/N ratio of non linear structures such as galaxy clusters, i.e
those structures which are most sensitive to $f_\mathrm{NL}$,
\begin{equation}  \label{eq:optimal}
  \hat Q(\boldsymbol\ell) = \alpha
  \frac{\hat\tau(\boldsymbol\ell)}{P(\ell)}, \quad\mbox{with}\quad
  \alpha^{-1}=\int_{\mathbb{R}^2} d^2\boldsymbol\ell
  \frac{\left|\hat\tau (\boldsymbol\ell)\right|^2}{P(\ell)} \;.
\end{equation}
Here the filter is represented in the Fourier space for convenience,
$\hat\tau(\boldsymbol\ell)$ is the Fourier transform of the expected
shear profile of individual halos, in our case the Fourier transform
of the NFW shear profile, and
$P(\ell)=P_{\gamma_\mathrm{t}}(\ell)+P_{\rm \epsilon}(\ell)$ described
in Section \ref{sec:cosmoConstraints}.  Note that the shape of the
optimal filter depends on the assumed parameters for the NFW density
profile, particularly on the angular size corresponding to the scale
radius $\theta_\mathrm{s}$. Moreover, since the filter is only a
radial function, its Fourier transform depends only on $\ell
=\|\boldsymbol{\ell}\|$

We further define the variance of the aperture mass estimate $A$ as
\begin{equation}
  \label{eq:Avariance}
  \sigma_A^2\equiv\int_{0}^{+\infty}\frac{\ell
    d\ell}{2\pi}P_\epsilon(\ell)|\hat{Q}(\ell)|^2 \;,
\end{equation}
so that it is related only to the non-cosmological signal component,
i.e. $P_\epsilon$. It is worth mentioning that in practical
applications $A$ is approximated as
\begin{equation}
  A(\boldsymbol\theta)=\frac{1}{n}\sum_{i=1}^n\epsilon_{{\rm
      t},i}(\boldsymbol{\theta})
  Q(\|\boldsymbol{\theta}_i-\boldsymbol{\theta}\|) \;,
  \label{eq:Aest}
\end{equation}
where $\epsilon_{{\rm t},i}(\boldsymbol{\theta})$ is the tangential
ellipticity of a galaxy image located at the position
$\boldsymbol\theta_i$ with respect to $\boldsymbol\theta$, and which
provides an estimate for $\gamma_{\rm t}$.

\subsection{Expected shear peaks number counts}\label{sec:expectedPeaks}

Once the lensing signals and the observational strategy have been
defined (see Sections~\ref{sec:cosmoConstraints} and
\ref{sec:filter}), the weak lensing number counts can be predicted as
the sum of two independent components. On one hand the non-linear
structures, i.e. galaxy clusters, whose detection depends on their
intrinsic abundance given by the mass function discussed in
Section~\ref{sct:mf}, and by their observed S/N ratio depending on
their expected signal $A$ (see Eq. \ref{eq:A}) and variance
$\sigma^2_A$ (see Eq. \ref{eq:Avariance}). On the other hand the
lensing LSS and instrumental noise which are well approximated by a
Gaussian random field and for which their number counts above a
specific threshold S/N$_\mathrm{th} = y_\mathrm{th}$ can be easily
predicted,
\begin{equation}
  \label{eq:detections}
  n_\mathrm{det}(y_\mathrm{th})=\frac{1}{4\sqrt{2}\pi^{3/2}}
  \left(\frac{\sigma_1}{\sigma_A}\right)^2
  \frac{y_\mathrm{th}}{\sigma_A}
  \exp\left(-\frac{y_\mathrm{th}^2}{2\sigma_A^2} \right) \;,
\end{equation}
as shown by \cite{MA09.1}. Here $\sigma_1$, defined as
\begin{equation}
  \label{eq:specMom}
  \sigma_1^2=\int_{0}^{+\infty}\frac{\ell^{3}d\ell}{2\pi}P(\ell)|\hat{Q}(\ell)|^2
  \;,
\end{equation}
is the LSS plus noise lensing field variance which depends on the
observational noise, the convergence power spectrum and the adopted
filter.
\begin{figure*}
  \includegraphics[width=0.8\hsize]{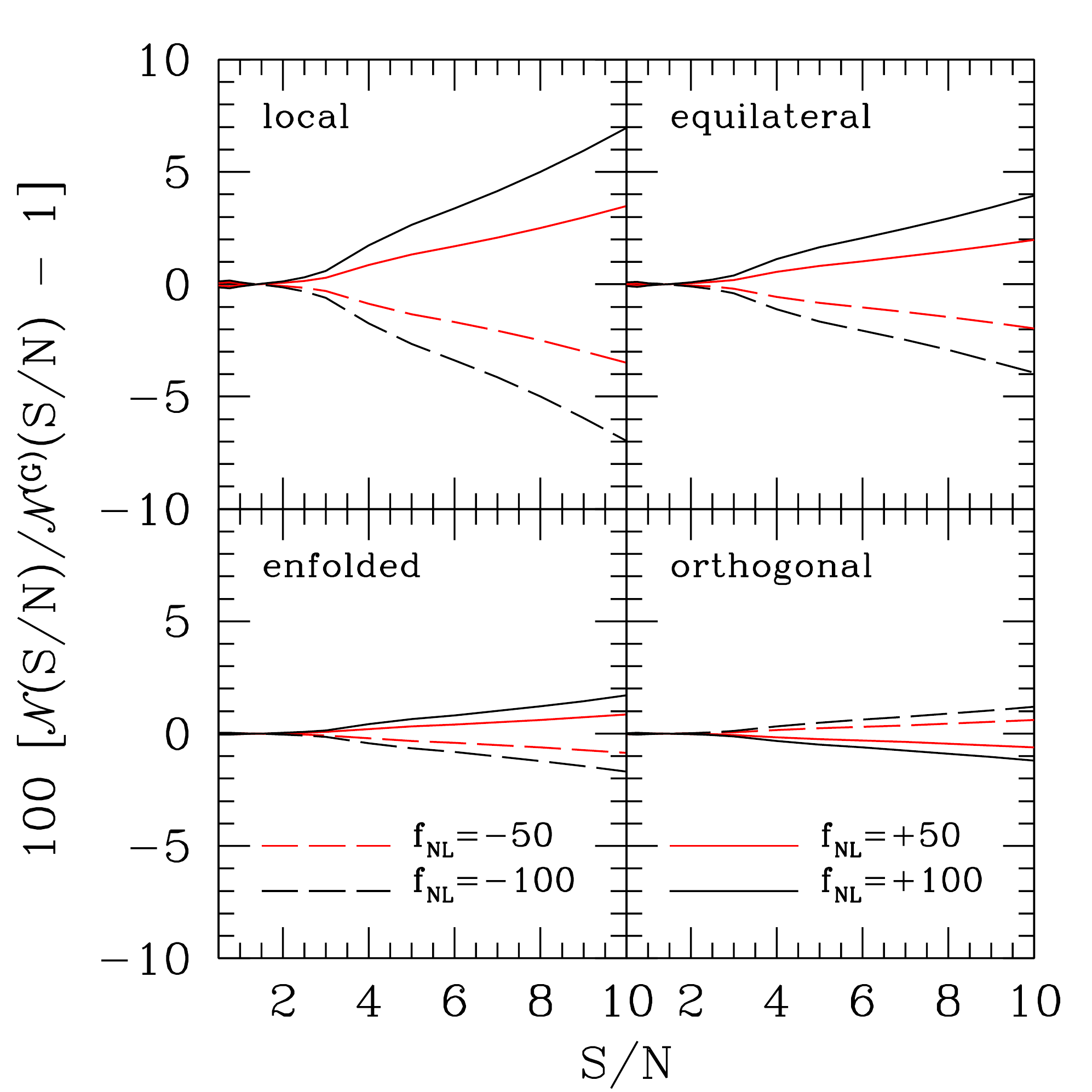}\hfill
  \caption{The ratio of the peak number counts in different
    models with primordial non-Gaussianity to the same
    quantities in the reference Gaussian $\Lambda$CDM cosmology, as a
    function of the S/N ratio. Four different non-Gaussian
    shapes and four different values of $f_\mathrm{NL}$ are
    shown, as labeled in the plot.}
  \label{fig:shearPeaksRatio}
\end{figure*}
Note that, by approximating this therm as a Gaussian random field, the
primordial non-Gaussianity is accounted through the LSS power spectrum
only so that its leverage is not fully included. For our purpose this
is a negligible approximation since we anyway expect this effect to be
small and put our final predictions on a conservative side.  These two
contributions carry complementary cosmological information, thus the
statistics of cosmic shear peaks are in principle valuable
cosmological tools \citep{DI09.1}.

The two cosmological contributions to the abundance of cosmic shear
peaks are exemplified in the left panel of Figure~\ref{fig:shearPeaks}
for the reference Gaussian cosmological model. For this Figure and in
the rest of the work we adopted the source redshift distribution
observationally derived by \citet{BE07.2}, while their average surface
number density is $n_\mathrm{g} = 40$ arcmin$^{-2}$, a level achieved
by actual weak lensing surveys and which should be reached by EUCLID
(a prediction for a sample of existing and future weak lensing surveys
is given by \citealt{MA09.1}). As expected, the chance alignment of
the LSS dominates the counts at small S/N ratio values, while the
contribution coming from the occurrence of real dark matter clumps
starts to be relevant at around S/N$\sim 4$ and dominates at higher
S/N values. We additionally show in the right panel of the same Figure
the total number counts expected for three different scale radii of
the optimal filter. As it can be noted, the larger the scale, the
larger the contribution of galaxy clusters with respect to the LSS and
the noise, although the latter remains always dominant at smaller, but
still large, S/N. This behavior is expected because for larger filter
scales the detections will have larger areas, thus increasing their
blending where their number density is higher, typically at lower
S/N. Also, the overall S/N ratio values at higher S/N are expected to
be larger, since a larger filter scale implies a better statistic for
the background galaxies reducing their shot noise and intrinsic
ellipticity. These features are ensured by the adopted optimal filter
which maximizes the cluster signal against the one of LSS and
noise. In the remainder of this paper we shall refer mainly to a scale
radius of $2'$, unless explicitly noted otherwise.

\section{Signature of primordial non-Gaussianity}\label{sct:effect}

The effect of primordial non-Gaussianity on the statistics of shear
peaks is exemplified in Figure \ref{fig:shearPeaksRatio} for the four
different non-Gaussian shapes that were discussed in
Section~\ref{sct:ng}, and for levels of non-Gaussianity
$f_\mathrm{NL}=\pm 50$ and $f_\mathrm{NL}=\pm 100$. As it can be seen,
the effect of non-Gaussianity is more evident at high values of S/N
ratio, where it is mostly given by the effect on the mass function of
dark matter halos. We remind the reader that our predictions regarding
the impact of primordial non-Gaussianity at the S/N ratios dominated
by the LSS are conservative as discussed in
Section~\ref{sec:expectedPeaks}. Also, as expected, the effect is
largest for local non-Gaussianity for which, at high S/N, the counts
can be modified by up to $\sim 7\%$ for $f_\mathrm{NL}=\pm100$ and
smallest for the orthogonal shape, where it barely reaches $\sim
1\%$. For the orthogonal shape positive values of $f_\mathrm{NL}$
provide a decrement in the number counts of shear peaks at high S/N
values, while the opposite is true for all other models. This agrees
with the behavior of the skewness that has been discussed in Section
\ref{sct:ng}. Although the effect of non-Gaussianity is substantially
reduced for shapes different from the local one, it is likely that
useful constraints can be put on $f_\mathrm{NL}$ for these shapes as
well. For instance in the orthogonal case, $f_\mathrm{NL}$ is
constrained only at the level of a few hundreds by CMB data
\citep{KO10.1}. Our prediction is compatible with the recent work of
\citet{MA10.2}, who found an effect on the shear peaks number counts
at the level of $\sim 10\%$ for the highest significance
detections. Although a direct comparison of the two works is not
possible due to different source redshift distributions, different
detection schemes, etc., we actually verified that by reducing the
scale radius of the optimal filter down to $\theta_\mathrm{s}=1'$ the
effect is raised to $\sim 10\%$.  Thus, we conclude that the two works
are in broad agreement for the local non-Gaussian model concerning the
magnitude of the effect. This is reassuring, since the two results are
based on quite different premises and adopt different methodologies:
we opted for semi-analytic modeling, while \citet{MA10.2} resorted to
the use of cosmological simulations.

\begin{figure}
	\includegraphics[width=\hsize]{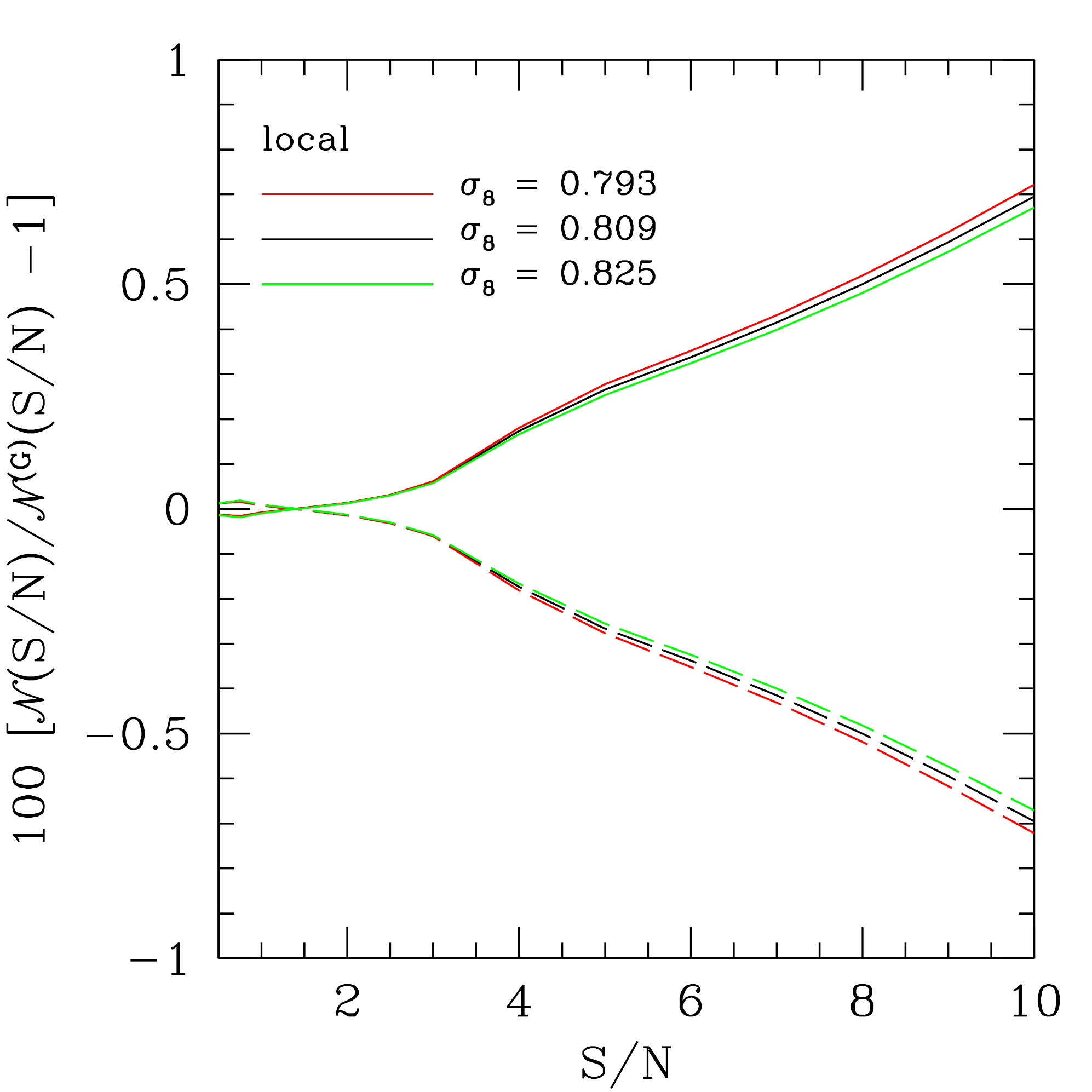}\hfill
	\caption{The ratio of the peak number counts in two different
          non-Gaussian models with local bispectrum shape to the same
          quantities in the reference Gaussian $\Lambda$CDM cosmology, as a
          function of the S/N ratio. The upper set of curves refer to
          $f_\mathrm{NL}=10$, while the lower set refers to
          $f_\mathrm{NL}=-10$. Different colors refer to different
          values of the matter power spectrum normalization
          $\sigma_8$, as labeled.}
	\label{fig:shearPeaksRatio_sigma8}
\end{figure}

In Figure \ref{fig:shearPeaksRatio_sigma8} we show the ratio of the
peak number counts in non-Gaussian cosmologies with local shape
($f_\mathrm{NL}=\pm 10$) to the same quantities in the reference
Gaussian case adopting three different values of the power spectrum
amplitude $\sigma_8$ for all models. We considered $\sim 2\%$
fluctuations in the value of $\sigma_8$, comparable with the
$1-\sigma$ uncertainty around the fiducial value derived by the WMAP-7
release. Quite unexpectedly, the effect of primordial non-Gaussianity
is higher for lower values of $\sigma_8$. We interpret this fact as
due to the delayed structure formation implied by a lower
normalization. This allows the effect of primordial non-Gaussianity on
the matter density field to be retained for a longer time by the LSS,
with the consequence that the S/N peak number counts are slightly more
affected. In other words, since with a lower $\sigma_8$ the onset of
non-linear evolution is delayed, gravitational clustering has less
time to erase the cosmological initial conditions.

\section {Cosmological constraints}\label{sct:constraints}

In this section we use a Fisher-matrix analysis to estimate the
confidence level in the $\sigma_8 - f_\mathrm{NL}$ plane, as obtained
with the weak lensing number counts.

\subsection {Fisher matrix analysis} \label{sec:fisher}

The Fisher matrix is given by
\begin{equation}
  F_{ij}=-\Bigg\langle\frac{\partial^2 \mathcal{L}}{\partial\xi_{i}
    \partial\xi_{j}}\Bigg\rangle \;,
\end{equation}
where $\mathcal{L}$ is the logarithm of the likelihood function and
$\boldsymbol{\xi}=(\sigma_8,f_\mathrm{NL})$ are the free parameters of
the number counts model. In case of a multivariate Gaussian
likelihood, the Fisher matrix can be written as
\begin{equation} \label{eq:Fisher}
  F_{ij}=\frac{1}{2}\mathrm{Tr}\left[A_{i}A_{j}+C^{-1}M_{ij}\right] \;,
\end{equation}
where $A_i=C^{-1}C_{,i}$, $M_{ij}=2(\partial \mu/\partial \xi_i)
(\partial \mu/\partial \xi_j)$, $C$ is the data covariance matrix, and
$\mu$ is the assumed model, i.e. the peaks number counts for different
S/N. Here a comma denotes differentiation with respect to the relevant
cosmological parameter. Since $C$ has a very weak dependence on the
model parameters $\sigma_8$ and $f_\mathrm{NL}$ the first term in
Eq.~\ref{eq:Fisher} is negligible because of $C_{,i}$. We evaluated
the Fisher matrix at the fiducial point $\sigma_8=0.809$ and
$f_\mathrm{NL}=0$ as measured by WMAP-7 for a standard $\Lambda$CDM
model \citep{KO10.1}.

As discussed in Section \ref{sec:expectedPeaks}, the number counts
model $\mu$ is given by two components. On one hand, the non linear
structures whose number counts are independent for different S/N ratio
bins and whose covariance contribution is therefore diagonal with a
Poissonian amplitude. On the other hand, the contribution given by the
LSS and noise for which the number counts are relevant for low S/N
ratios up to S/N $\sim 3$, where we expect to have a small but not
negligible covariance contribution since different S/N bins can be
correlated. To account for the non-diagonal covariance elements, we
realized a set of 1000 numerical realizations of weak lensing data,
modeled with a Gaussian random field whose power spectrum, $P(\ell)$,
is discussed in Section \ref{sec:cosmoConstraints}. As expected the
number counts derived from the simulation agrees with the analytical
predictions except for very low values of the S/N ratio (S/N $<0.2$)
where the analytic approximation fails as discussed by
\cite{MA09.1}. In any case these very low S/N are completely
negligible for any cosmological analysis.

\begin{figure}
  \centering
  \includegraphics[width=1.2\hsize, angle=-90]{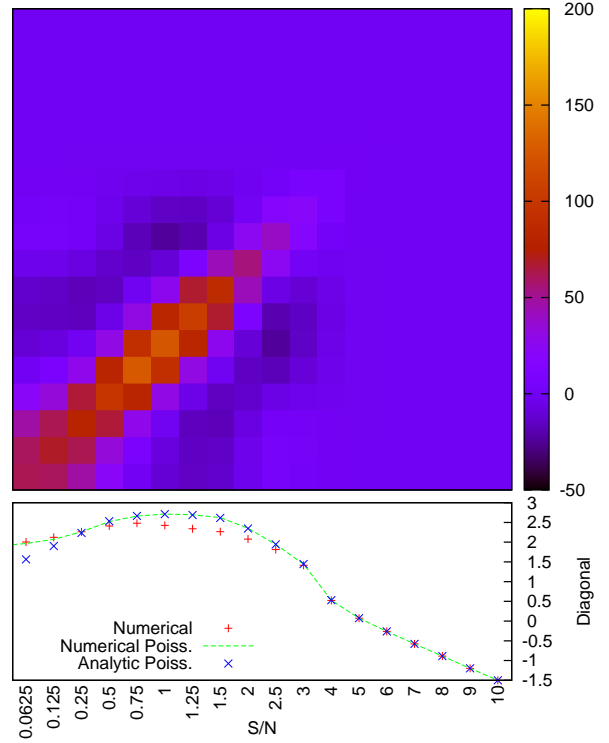}
  \caption{Weak lensing number counts covariance matrix for a
    $\Lambda$CDM model. The off-diagonal terms are small but not
    negligible only at low S/N ratios. The bottom panel shows the
    comparison between the diagonal components of the full covariance
    matrix derived numerically (shown in the top panel) and what would
    be expected for ideal number counts with Poisson statistic and
    statistically independent S/N bins. For high S/N the covariance is
    diagonal and Poissonian by construction.}
  \label{fig:covariance}
\end{figure}

The total covariance matrix of weak lensing number counts is shown in
Figure \ref{fig:covariance}. As expected, it has a strong diagonal
component and its off-diagonal terms are small but non negligible only
at low S/N ratios. By treating the two components, LSS + noise and
galaxy clusters as independent, we do not account for their mutual
influence on the final number counts, but this effect is expected to
be small because the signature of each cluster affects only few
pixels. In any case, this approach is conservative with respect to the
final cosmological constraints because we neglect the impact of the
detections with high S/N ratios, carriers of the non-Gaussianity
signal, on those with lower S/N ratios.
    
\subsection {Cosmological constraints on $f_\mathrm{NL}$}

\begin{figure}
  \centering
  \includegraphics[width=\hsize] {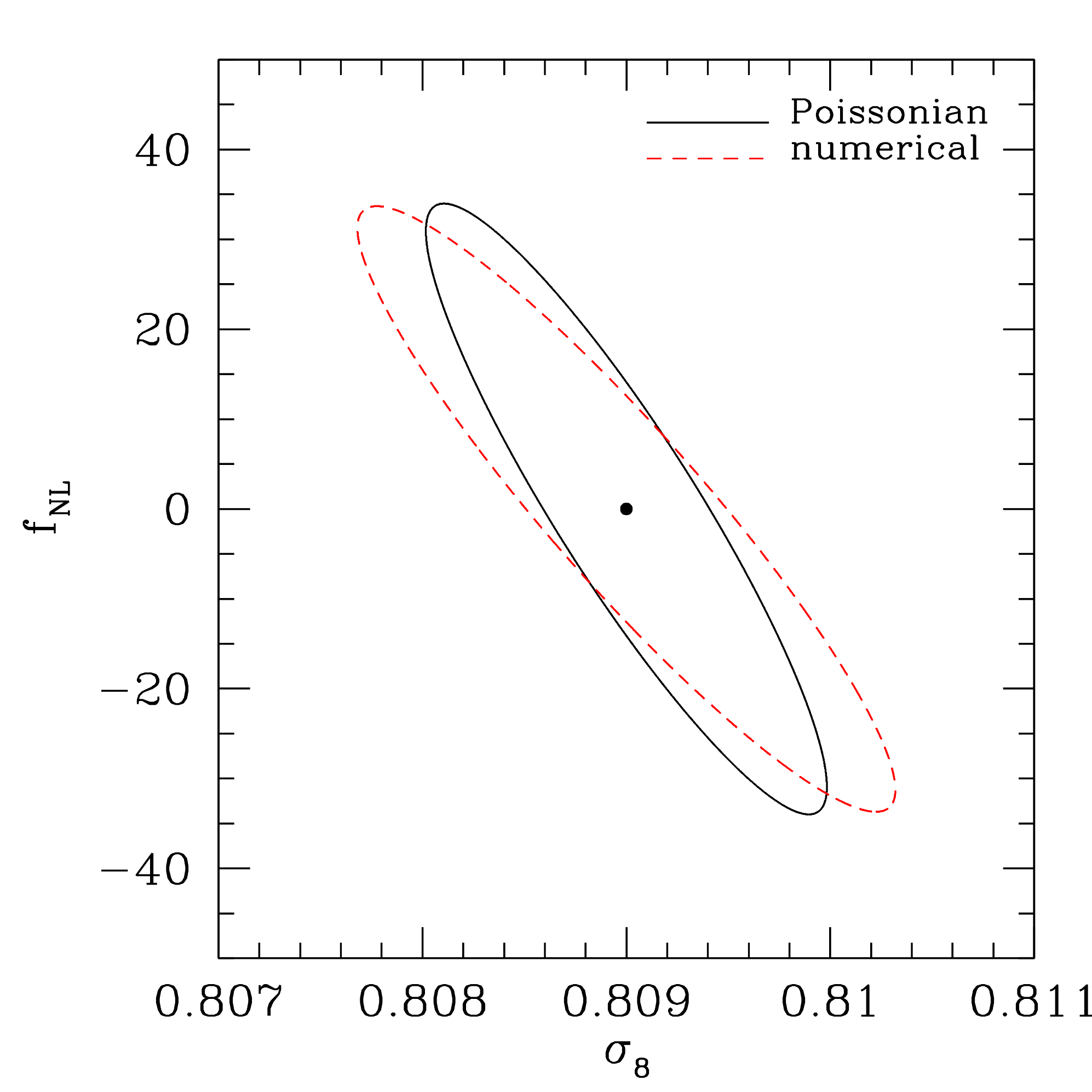}
  \caption{Cosmological $1-\sigma$ confidence levels as obtained for
    the local shape non-Gaussian model using weak lensing number
    counts, by assuming both Poissonian statistical independence
    (Poissonian), and by using the numerical simulations including the
    full covariance matrix (numerical).}
  \label{fig:comparisonSimAnal}
\end{figure}

In this Section we assume the expected performances of the weak
lensing survey \emph{Euclid} \citep{LA09.1}, i.e. a field of view of
$20,000$ deg$^2$, an average background galaxy number density of
$n_\mathrm{g}=40$ arcmin$^{-2}$ with an intrinsic ellipticity
\emph{rms} of $\sigma_\mathrm{\epsilon_s}=0.3$, and a redshift
distribution similar to the one discussed by \cite{BE07.2}. The
confidence levels are evaluated with the Fisher matrix analysis
discussed in Section \ref{sec:fisher}. In Figure
\ref{fig:comparisonSimAnal} we compare the confidence regions for the
local shape non-Gaussian model obtained first by assuming purely
Poissonian independent statistic, i.e. a data covariance having a pure
diagonal component with Poissonian amplitude, and then by using the
full covariance matrix obtained from the numerical simulations (see
Section \ref{sec:fisher}). As can be seen, the $1-\sigma$ confidence
ellipse is slightly tilted in the latter case with respect to the
former, implying a somewhat different direction of degeneracy between
the two parameters under consideration. Overall, however, the
constraints on the non-Gaussianity level $f_\mathrm{NL}$ are basically
unchanged, while those on the normalization of the matter power
spectrum $\sigma_8$ are only very slightly loosened by using the
simpler Poisson model. As a consequence, it is safe to state that the
final differences between the two approaches are negligible and that
for a first fast analysis the simplified approach can be used.

\begin{table}
  \label{tab:cramerRao}
  \centering
  \begin{tabular}{c|cc}
    \hline
    \hline
    $n_\mathrm{g}$ & $\Delta \sigma_8$ & $\Delta f_\mathrm{NL}$\\
    \hline
    20 & 0.0023 & 60\\
    40 & 0.0015 & 45\\
    80 & 0.0012 & 41\\
    \hline
    \hline
  \end{tabular}
  \begin{tabular}{c|cc}
    \hline
    \hline
    $\theta_\mathrm{s}$& $\Delta \sigma_8$ & $\Delta f_\mathrm{NL}$\\
    \hline
    1' & 0.0016 & 57\\
    2' & 0.0015 & 45\\
    4' & 0.0015 & 38\\
    \hline
    \hline
  \end{tabular}
  \caption {Dependence of the $1-\sigma$ joint bounds on the
    background galaxy number density $n_\mathrm{g}$ (left) and on the
    filter scale radius, $\theta_\mathrm{s}$ (right). The fiducial
    model with $\sigma_8=0.809$ and $f_\mathrm{NL}=0$ is adopted.}
\end{table}

Given that, we used the first approach to estimate the Cram\'er-Rao
bounds on $\sigma_8$ and $f_\mathrm{NL}$ at $1-\sigma$ level for
different survey depths, i.e. different galaxies number densities, and
for different filter scales as shown in Table 1. In order to save
computational time, the full data covariance has been used only for
the final results reported below. The results are evaluated for a
survey sky coverage of $20,000$ deg$^2$ and scale with the square root
of the survey area. The constraints on both $f_\mathrm{NL}$ and
$\sigma_8$ improve with increasing number density of background
sources. However only a marginal benefit is obtained by increasing the
number density over $40$ arcmin$^{-2}$. In other words, in terms of
joint constraints for the level of non-Gaussianity and amplitude of
the matter power spectrum, it does not pay off to increase the depth
of the survey beyond what is already planned for
\emph{Euclid}. Concerning the scale radius of the optimal filter, by
increasing it we get a significant tightening of the constraints on
$f_\mathrm{NL}$, while those on $\sigma_8$ are almost insensitive to
it.

\begin{figure}
  \centering
  \includegraphics[width=\hsize]{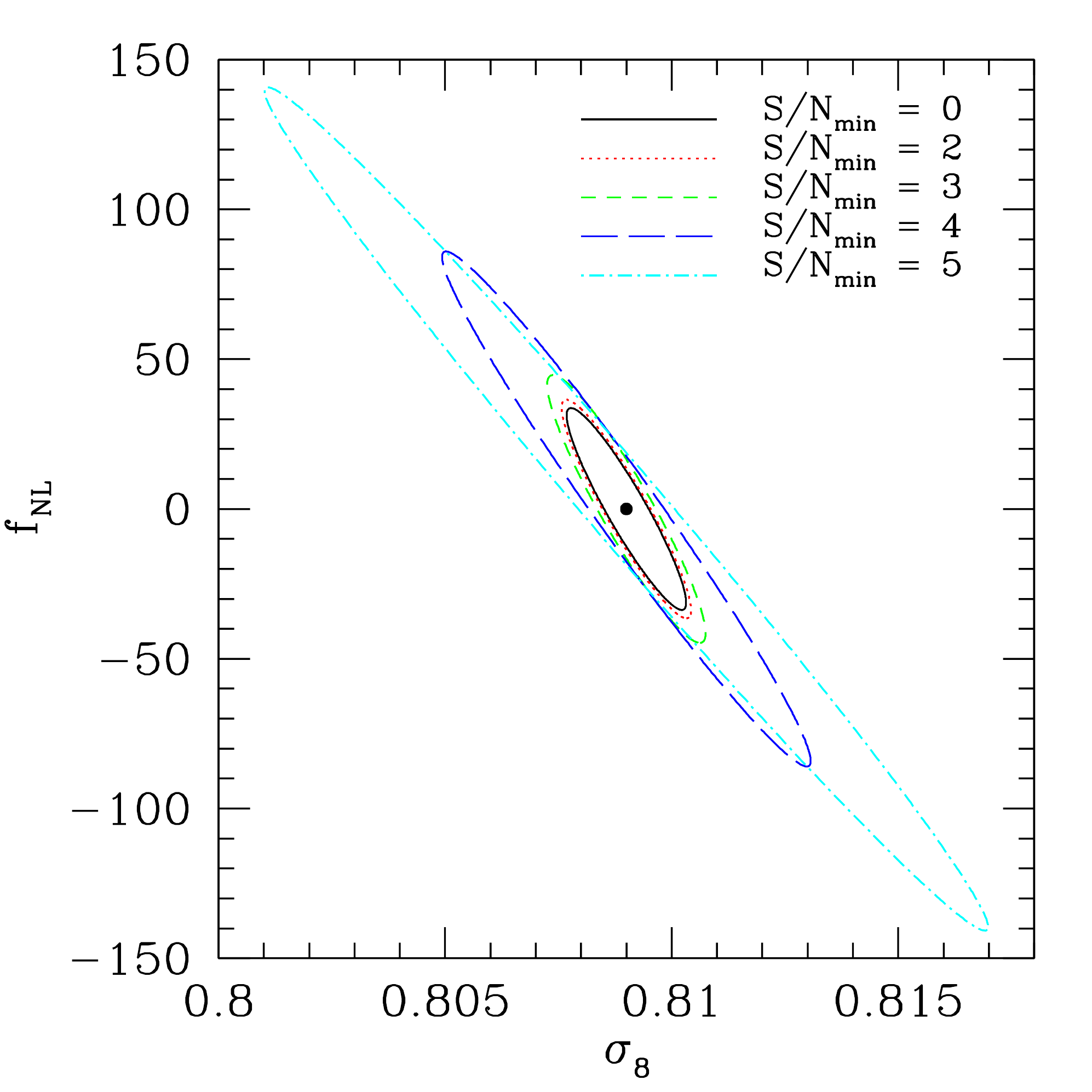}
  \caption{Confidence contours for the local shape non-Gaussian model,
    with different lower S/N cut-off limits. This shows the
    information lost by cutting the data at different S/N ratios.}
  \label{fig:ellipse-sn}
\end{figure}

\begin{figure*}
  \centering
  \includegraphics[width=0.8\hsize]{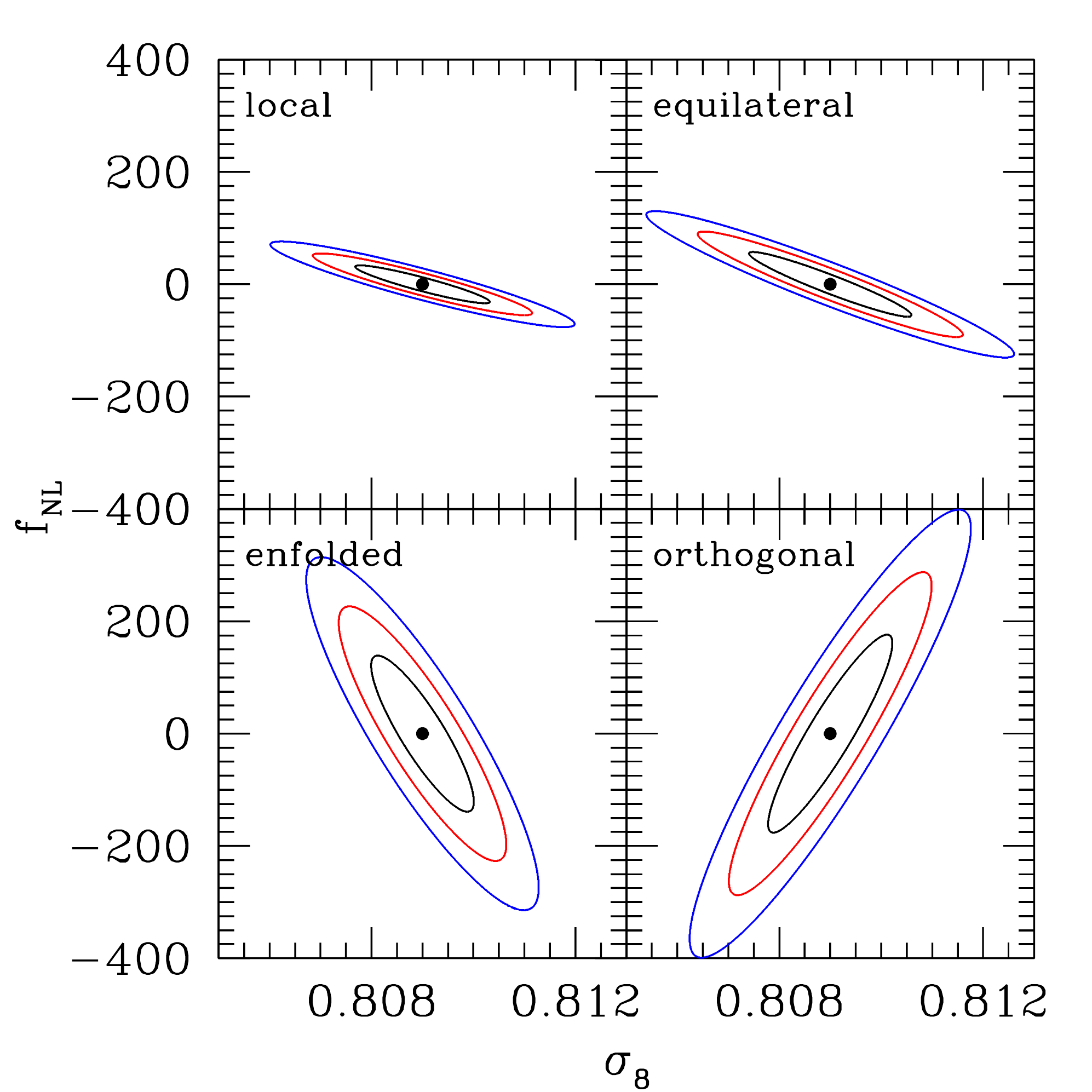}
  \caption{Confidence regions for the four adopted models with
    non-Gaussian initial conditions. We report results for the local
    (top left), equilateral (top right), enfolded (bottom left) anqd
    orthogonal (bottom right) primordial bispectrum shapes. The three
    contours in each panel refer to $68.3\%$, $95.4\%$, and $99.7\%$
    confidence levels.}
  \label{fig:ellipseModels}
\end{figure*}

Now that we have an overall picture of the dependence on the survey
characteristics, we used the numerical simulations including the full
covariance matrix to obtain our final results. In
Figure~\ref{fig:ellipse-sn} we show the $1-\sigma$ confidence levels
in the $\sigma_8-f_\mathrm{NL}$ plane obtained for the non-Gaussian
model with local shape when different minimum S/N ratios are
considered. It is in fact common practice for actual applications to
use only those detections with S/N larger then S/N$_\mathrm{min}=3-5$
in the attempt to isolate galaxy clusters from all other signal
components. Here we show how the cut-off level affects the final
results. The related loss of information happens because the presence
of non linear structures affects the statistic of weak lensing maps
also at lower signal to noise ratios where they cannot be seen as
clear single detections but where their presence is still visible.
Moreover, although this effect is negligible for the $f_\mathrm{NL}$
estimates, chance projection of the LSS dominates at these small S/N
values, and still contains cosmological information. Figure
\ref{fig:ellipse-sn} shows that adopting a conservative choice for the
minimum S/N value can results in a loss of constraining power of a
factor $\sim 3$ on $f_\mathrm{NL}$ and a factor of $\sim 4$ on
$\sigma_8$. Thus, particular care have to be used when defining the
detection selection criteria since data can be fully exploited only if
lower S/N levels, e.g. S/N$_\mathrm{min}\sim 2$, are used.  This would
be possible only if a deep statistical understanding of data is
achieved together with a detailed modeling. A tentative to go in this
direction is provided by the analytic prediction for weak lensing
number counts proposed by \cite{MA09.1} but which still needs to
improve some of the adopted approximations.

Finally we show in Figures \ref{fig:ellipseModels} the constraints for
$68.3\%$, $95.4\%$ and $99.7\%$ confidence levels with
S/N$_\mathrm{min}=0$ for the four different primordial bispectrum
shapes discussed in Section \ref{sct:ng}. We adopted the optimal
filter with scale of $\theta_\mathrm{s}=2'$. As one could naively
expect, while the constraints on $\sigma_8$ are rather insensitive to
the adopted shape of the primordial bispectrum, the constraints on
$f_\mathrm{NL}$ vary widely with it. Particularly, while the
$1-\sigma$ error $\Delta f_\mathrm{NL}$ is at the level of a few tens
for the local and equilateral shapes, it grows up to $\sim 100-200$
for the enfolded and orthogonal shapes. These constraints are not
competitive with what is expected by future CMB and galaxy clustering
probes, and are only comparable with the expected performance of the
weak lensing power spectrum \citep{FE10.1} analysis. The main reason
for this results is that in this work we did not use the full redshift
information which is otherwise included in the other works and which
could be included by the use of lensing tomography. Recently,
\citet{FE10.2} have shown that using the correlation function of weak
lensing-selected clusters it is possible to push the bounds on
$f_\mathrm{NL}$ below what would be achieved without redshift
information. In addition, it is likely that a combination of these
diverse probes can reduce the error on $f_\mathrm{NL}$
substantially. We plan to investigate this synergy in a future work.

The constraints on the amplitude of the matter power spectrum, $\Delta
\sigma_8 \sim 10^{-3}$ are themselves quite competitive. For instance,
recently \citet{WA10.2} forecasted an error on $\sigma_8$ of $\sim
7\times 10^{-2}$ by using the shape and position of the Baryon
Acoustic Oscillation measured by \emph{Euclid}. Despite the fact that
\citet{WA10.2} let all the cosmological parameters free to vary, while
we limit our analysis to $f_\mathrm{NL}$ and $\sigma_8$, these numbers
argue for our constraints on the amplitude of the matter power
spectrum being sound.

\begin{figure}
  \centering
  \includegraphics[width=\hsize] {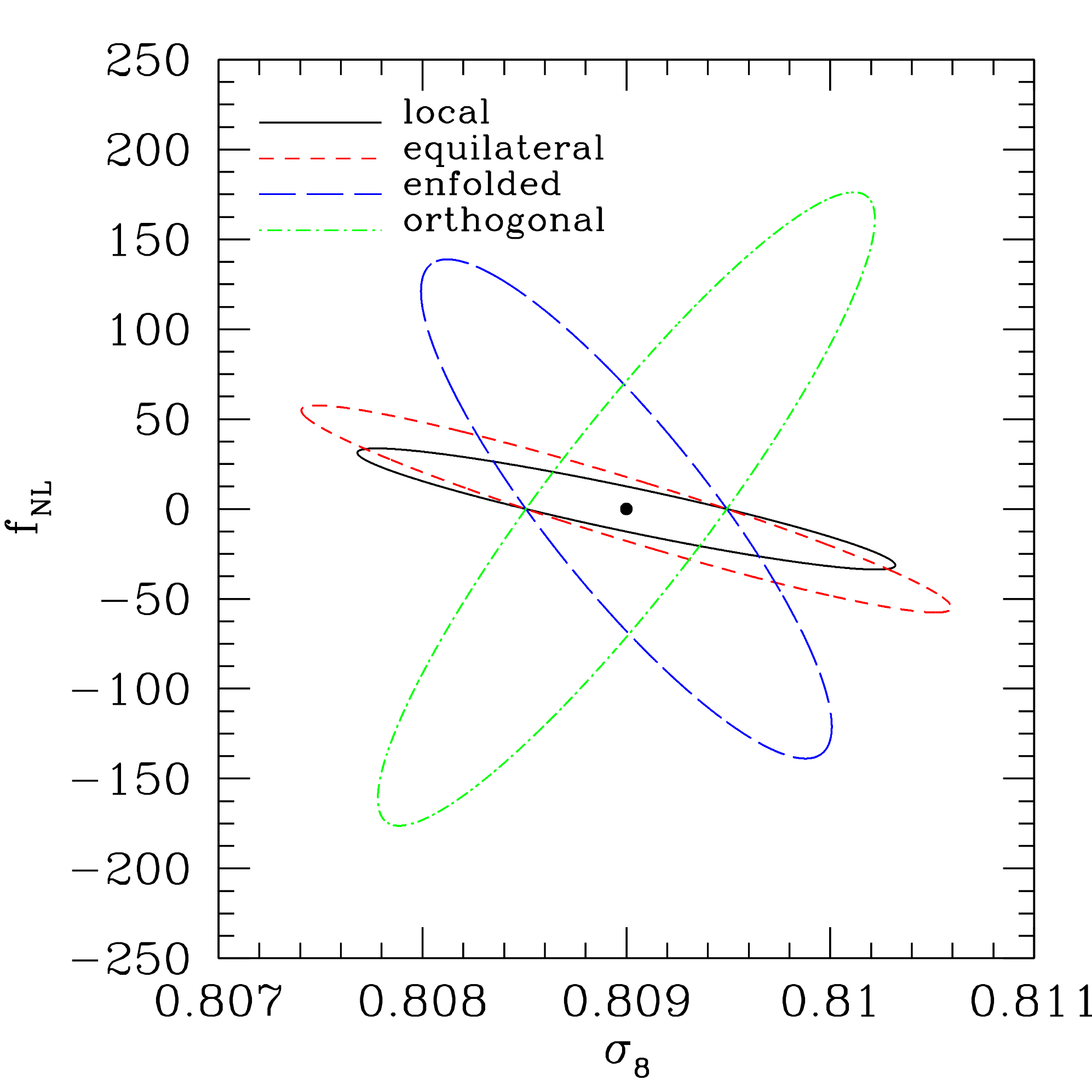}
  \caption{The $1-\sigma$ confidence levels in the
    $\sigma_8-f_\mathrm{NL}$ plane for all the four shapes of the
    primordial bispectrum considered in this work, as labeled}
  \label{fig:compare}
\end{figure}

The degeneracy between the level of primordial non-Gaussianity and the
amplitude of the matter power spectrum is also easy to understand,
since increases in both $f_\mathrm{NL}$ and $\sigma_8$ bring an
increment in the large scale matter power spectrum and the occurrence
of massive dark matter halos. The only exception to this is given by
the orthogonal model, because in this case a positive $f_\mathrm{NL}$
implies a decrement in both the abundance of cosmic structures and the
large scale bias of dark matter halos. For further clarity, in Figure
\ref{fig:compare} we show the $1-\sigma$ confidence levels in the
$\sigma_8-f_\mathrm{NL}$ plane for the four primordial bispectrum
shapes considered in this work together. This Figure allows a more
direct comparison between the different models, and highlights the
differences in constraining the level of primordial non-Gaussianity
between them.

\section{Summary and conclusions}\label{sct:conclusions}

The abundance of S/N peaks in cosmic shear maps contains cosmological
information through the mass function of large dark matter clumps and
the projection of the large scale matter distribution. We considered
the effect of deviations from primordial Gaussianity on this
particular weak lensing statistic. Analytic modeling of the galaxy
cluster mass function, the large scale halo bias, and the fully
non-linear power spectrum of dark matter was implemented, where the
effect of primordial non-Gaussianity could be straightforwardly
introduced. We predicted analytically and through simple numerical
simulations the expected weak lensing peaks number counts and
covariance for different weak lensing filters. We then performed a
Fisher matrix analysis in order to forecast the constraints on the
level of primordial non-Gaussianity $f_\mathrm{NL}$ and the amplitude
of the matter power spectrum $\sigma_8$ that could be expected by
counting the cosmic shear peaks for different survey configurations,
i.e. depth and field of view, in particular for the half-sky maps
expected by the proposed ESA space mission \emph{Euclid}. Our
principal results can be summarized as follows.

\begin{itemize}
\item Primordial non-Gaussianity affects mainly the high S/N part of
  the shear peak number counts, where the signal is dominated by the
  occurrence of real cluster-sized dark matter halos. For the local
  primordial bispectrum shape and according to the adopted weak
  lensing filter the number counts can be modified by $\sim 3-4\%$ at
  S/N $\gtrsim 10$ if $|f_\mathrm{NL}| = 50$, and by up to $\sim 10\%$
  for $|f_\mathrm{NL}| =100$. For the enfolded and orthogonal
  bispectrum shapes, the modification reaches at most $\sim 1-2\%$ for
  $|f_\mathrm{NL}|=100$.

\item Generically an increment in $f_\mathrm{NL}$ corresponds to an
  increment in the number counts of cosmic shear peaks, and the other
  way round, compatibly with the effect of primordial non-Gaussianity
  on the cluster mass function. The exception to this is given by the
  orthogonal model, for which it is known that the skewness and
  scale-dependent bias correction are negative for positive
  $f_\mathrm{NL}$, and vice-versa.

\item We show that considering only cosmic shear peaks with S/N larger
  than $3-5$, as it is costume in actual applications, leads to a loss
  of cosmological information because the LSS and, more importantly,
  the low mass clusters contributions are ignored. In particular the
  latter is important in constraining $f_\mathrm{ NL}$ even if these
  structures cannot be detected individually. Thus, data can be better
  exploited if lower S/N levels, e.g. S/N$_\mathrm{min} \lesssim 2$,
  are used. For this achievement, a deep statistical understanding and
  a detailed data modeling are necessary.

\item In terms of joint constraints for $f_\mathrm{NL}$ and $\sigma_8$
  only a marginal benefit is obtained by increasing the background
  galaxy number density above $40$ arcmin$^{-2}$, while it is more
  convenient to go for wider fields of view. By increasing the scale
  radius of the optimal filter we obtained a significant improvement
  for the $f_\mathrm{NL}$ constraints, while those on $\sigma_8$ are
  almost unaffected.

\item Counting cosmic shear peaks in future wide field
  optical/near-infrared surveys on the model of \emph{Euclid} can
  constrain $f_\mathrm{NL}$ to the level of a few tens for the local
  and orthogonal shapes, and to the level of $\sim 100-200$ for the
  enfolded and orthogonal shapes. Constraints on $\sigma_8$ are at the
  level of $\sim 10^{-3}$ and are instead rather insensitive to the
  assumed shape of the primordial bispectrum.
\end{itemize}

The results presented in this work show that the abundance of cosmic
shear peaks can be a powerful mean for constraining
cosmology. Forecasted errors on the level of primordial
non-Gaussianity for a \emph{Euclid}-like survey are competitive with
similar constraints given by weak lensing tomography and the
correlation function of weak lensing selected galaxy clusters.  At the
same time, counting cosmic shear peaks irrespective of their nature is
much easier and less subject to systematics compared with other probes
where the nature of each lensing peak, resulting by LSS or galaxy
clusters, has to be determined. Our findings outline once more the
great advance in understanding the physics of the primordial Universe
that is expected thanks to future wide field imaging surveys.

\section*{Acknowledgments}

This work was supported in part by the University of Heidelberg and
the Transregio-Sonderforschungsbereich TR 33 of the Deutsche
Forschungsgemeinschaft. We acknowledge financial contributions from
contracts ASI-INAF I/023/05/0, ASI-INAF I/088/06/0, ASI I/016/07/0
'COFIS', ASI '\emph{Euclid}-DUNE' I/064/08/0, ASI-Uni
Bologna-Astronomy Dept. '\emph{Euclid}-NIS' I/039/10/0, and PRIN MIUR
'Dark energy and cosmology with large galaxy surveys'.

\end{document}